# Electronic Landscape of Group-5 Transition Metal Ditellurides $M\mathrm{Te}_2$ ($M$ = V, Nb, Ta): Multiple Crystal Phases with Local Bonds and Flat Bands

Natsuki Mitsuishi[1,2*] and Kyoko Ishizaka[1,2†]

[1]*RIKEN Center of Emergent Matter Science (CEMS), Wako, Saitama 351-0198, Japan*

[2]*Quantum-Phase Electronics Center & Department of Applied Physics, The University of Tokyo, Hongo, Tokyo 113-8656, Japan*

**Group-5 transition metal ditellurides $M\mathrm{Te}_2$ ($M$ = V, Nb, Ta) are unique $\mathrm{CdI}_2$-type layered materials that exhibit peculiar quasi-one-dimensional intralayer superstructures, known as ribbon-chains and butterfly-like clusters of $M$ atoms. In this review article, we attempt to systematically understand their electronic band structures based on our recent angle-resolved photoemission spectroscopy (ARPES) studies and first-principles calculations. We underscore the role of the localized molecular-like orbital bonds that form the anomalous flat bands in the momentum space, and demonstrate how they influence the Fermi surface anisotropy, phase stabilities of crystal structures, and nontrivial topological properties in some cases. We also elaborate the future prospects for novel quantum phenomena that can be realized in these materials.**

*natsuki.mitsuishi[at]riken.jp

†ishizaka[at]ap.t.u-tokyo.jp

## 1. Introduction

Molecular-like orbital bonding, analogues of dimers, polymers or clusters in organic molecules, is becoming one of the key concept for organizing complex inorganic compounds with versatile physical properties [1]. This concept has been widely used in strongly correlated insulators with charge/magnetic/orbital ordering (e.g., transition metal oxides [2] and spinel compounds [3]), but recently it has also become important in phenomena like charge density wave (CDW) [4] and lattice-collapse transitions [5] in various metallic systems including superconductors. One important example is the transition metal dichalcogenides $MX_2$ ($M$ = transition metal, $X$ = S, Se, Te), the van der Waals layered materials hosting rich quantum phenomena including CDW ordering [6–8]. CDW is basically described as a superperiodic wave-like modulation of the electron density accompanying the concomitant lattice distortion, induced by the Peierls instability of Fermi-surface nesting in low-dimensional systems [9]. On the other hand, some of the superperiodic lattice distortions in $MX_2$ has been argued in terms of molecular bonding of relevant orbitals [10]. Especially in the case of $CdI_2$-type phase (hereafter referred to as 1*T*) [Figs. 1(a,b)], Whangbo *et al.* [10] proposed a mechanism of metal-metal bonding that leads to a variety of superperiodic patterns depending on the number of *d* electrons [Fig. 1(c)]. For this materials family, the $MX_6$ octahedral configuration splits the metal *d* orbitals into the low-lying $t_{2g}$ ($d_{XY}$, $d_{YZ}$, $d_{ZX}$) and high-lying $e_g$ ($d_{X^2-Y^2}$, $d_{Z^2}$) manifolds. For simplicity, the effect of trigonal distortion is ignored and $O_h$ symmetry is assumed here. The above scenario can be interpreted based on the concept of the directional σ-bonding of $t_{2g}$ orbitals residing on the edge-sharing octahedra, which introduces the scheme of virtual (or hidden) one-dimensional Fermi surfaces and their nesting according to the *d* electron occupation [10].

Of particular interest in this review is the group-5 ditelluride $M$Te$_2$ ($M$ = V, Nb, Ta) revealing the ribbon chain (sometimes termed double zigzag chain) superstructure [see the middle panel in Fig. 1(c)] [11,12]. This superstructure can be regarded as the superposition of $M$-trimers among adjacent $d_{YZ}$, and $d_{ZX}$ orbitals, which can be energetically stabilized by forming the so-called "three-center two-electron" state [the right panel in Fig. 1(b)] [10,13,14]. Here, the number of *d*-electrons is assumed to be close to 4/3 due to the fairly strong charge transfer from Te to $M$ [15]. We note that the importance of Te-Te hybridization and $M$-Te charge transfer has been argued in tellurides ($M$Te$_2$) [15–18], which may make the picture of $M$-$M$ and/or Te-Te local bonding more plausible as compared to sulfides and selenides. Indeed, there have been similar arguments in various tellurides, such as $M$ zigzag chains in (Mo,W)Te$_2$ [19] and (Ta,Nb)IrTe$_4$ [20], Ir dimer in IrTe$_2$ [21], and Te dimer in AuTe$_2$ [22]. Here we should also note that the origin of ribbon chain formation has been also discussed based on a typical Fermi surface nesting scenario [23] and/or anisotropic electron-phonon coupling [23,24] from first-principles calculations, whereas its *M* dependence is not yet reported. Furthermore, $TaTe_2$ exhibits another superstructure with butterfly-like Ta clusters [25] at low temperature, which is not presented in Fig. 1(c). Thus a systematic *M*- and temperature-dependent investigation on electronic properties is necessary.

In this review, we focus on the group-5 tellurides $MTe_2$ ($M$ = V, Nb, Ta) and attempt to systematically understand their electronic band structures based on angle-resolved photoemission spectroscopy (ARPES) and first-principles calculations (Some data are cited from our recent publications [26,27]). We primarily demonstrate the characteristic changes in electronic band structures accompanying the formation of ribbon chains and butterfly-like clusters. Here it is important to grasp the correspondence between the chemical orbital bonding in real space and the band structure in momentum space. We also discuss their impact on the Fermi surface anisotropy, crystal structural phase stabilities, and nontrivial topological properties.

This review is organized as follows. In Sect. 2, we briefly review the superperiodic patterns and phase transitions reported in $MTe_2$ ($M$ = V, Nb, Ta) and their associated physical properties. Section 3 describes the characteristics of the electronic band structures in the undistorted 1*T* trigonal and ribbon-chain phases, as revealed in the pristine- and Ti-doped $VTe_2$. In Section 4 we turn our eyes on to the common ribbon-chain phase in room temperature $MTe_2$ ($M$ = V, Nb, Ta), focusing on the *M*-dependent band structure and its phase instability. Section 5 overviews the temperature dependent electronic structure of $TaTe_2$ with a particular focus on its modifications through the ribbon-chain to butterfly-like-cluster phase transition. Section 6 summarizes the review and offers some suggestions for future research.

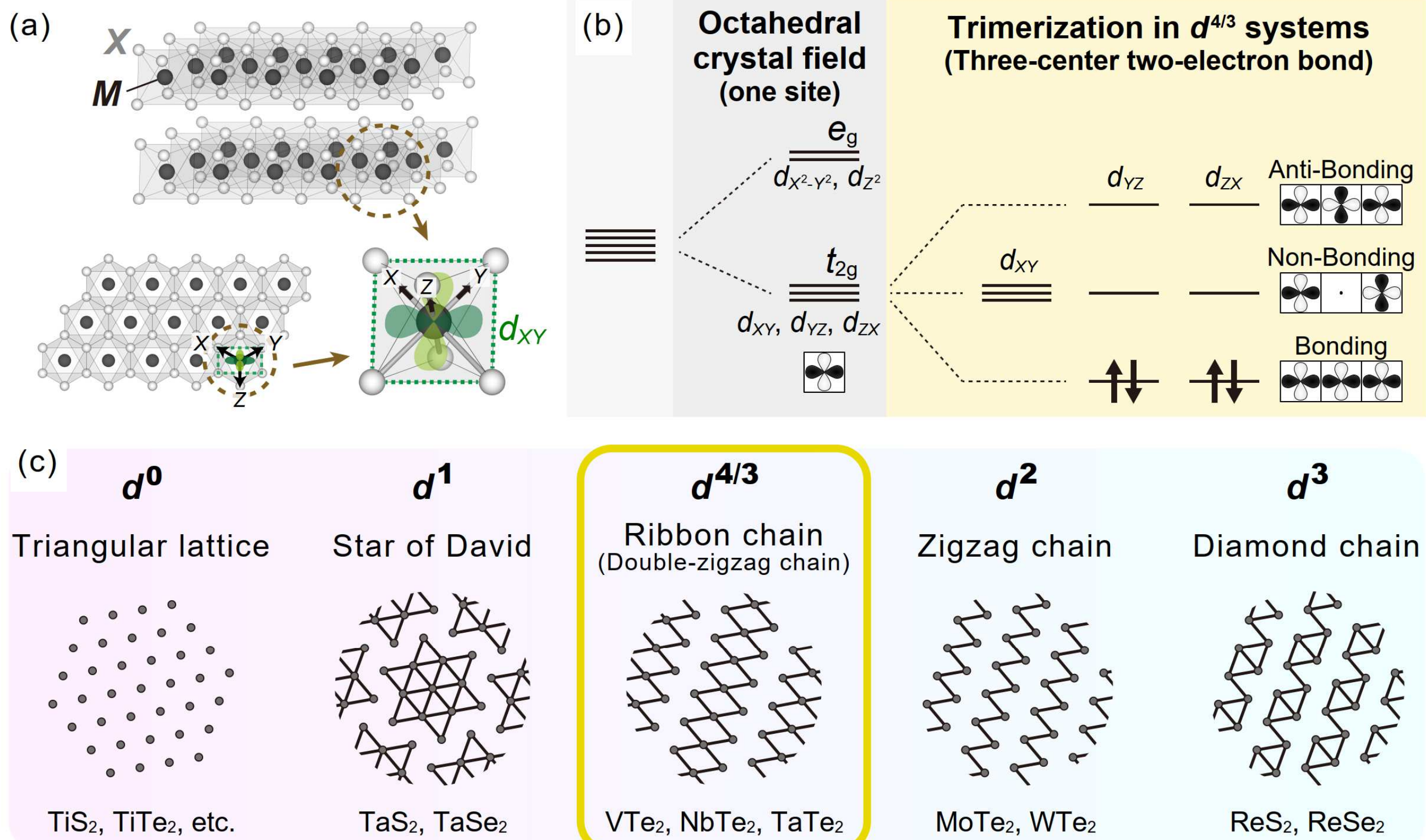

**Fig. 1. Crystal structure and metal-metal bonding in $CdI_2$-type transition metal dichalcogenides.** (a) Crystal structure of $MX_2$ with the edge-sharing $MX_6$ octahedral configuration. The crystal structures are visualized by VESTA [28]. $d_{XY}$ orbital is depicted in the octahedron. (b) Metal $d$ orbital level splitting under the octahedral crystal field [middle panel] and the trimerized "three-center two-electron" bonded state [right panel]. (c) Multiple metal-metal bonding representing the patterns of lattice distortions observed in various $t_{2g}$ systems [10]. This review focuses on group-5 $M\mathrm{Te}_2$ ($M$ = V, Nb, Ta) highlighted by the yellow box.

## 2. Superperiodic Patterns and Phase Transitions in *M*Te₂ (*M* = V, Nb, Ta)

Prior to elaborating the electronic band structures of the bulk group-5 $MTe_2$ ($M$ = V, Nb, Ta), we first summarize their superperiodic patterns and phase transitions as presented in Figs. 2(a)–(d). Among them, only $VTe_2$ crystallizes in the non-distorted 1*T* structure [Fig. 2(a), $P\bar{3}m1$] at high-temperature regime. $VTe_2$ undergoes a first-order structural phase transition to the monoclinic 1*T*" phase [Fig. 2(b), $C2/m$] at $T_s$ ~ 480 K upon cooling, appearing as a jump in the temperature-dependent resistivity [29]. The 1*T*" structure is characterized by the quasi-one-dimensional ribbon chains (double-zigzag chains) of V atoms stemming from the substantial contraction of the V-V bond distance of about 9% (~0.33 Å) [12] through the transition. This 1*T*" state is also referred as the metallic CDW state with (3×1×3) superstructure, where the last term "3" arises from the successive 1/3 shift of ribbon chains between adjacent layers [30]. Since $TiTe_2$ has the 1*T* structure down to the lowest temperature [31], the 1*T*–1*T*" transition temperature of $VTe_2$ can be suppressed by Ti doping as displayed in Fig. 2(e) [26]. We also note that the phase transition in $VTe_2$ can be feasibly manipulated by optical pulses as reported by recent pump-probe diffraction studies [30,32,33] and optical spectroscopy [34,35].

In contrast to $VTe_2$, the 1*T*" phases in $NbTe_2$ and $TaTe_2$ are thermally stable and persist up to the highest temperature (as will be elaborated in Sect. 4.1). Among them, only $NbTe_2$ maintains the 1*T*" structure in the whole temperature region and exhibits superconductivity below 0.5–0.7 K [36,37]. Meanwhile, $TaTe_2$ exhibits a first-order structural phase transition at $T_s$ ~ 170 K to a low-temperature (3×3×3) superstructure [Fig. 2(c), $C2/m$] [25]. This superstructure is distinguished by a large displacement of some Ta atoms along the $\mathbf{b}_m$-axis (up to ~0.28 Å), resulting in the formation of the "butterfly-like" clusters (or "heptamer" [38]) with multiple isolated Ta sites [25,39]. Several anomalies in transport properties are observed at $T_s$, including a step-like decrease in electrical resistivity (upon cooling) [25] and a change in the sign of the Seebeck coefficient [40], as shown in Fig. 2(f) [27]. Recent studies have demonstrated that this phase transition can be controlled through a variety of methods, including chemical doping [24,40–42], pressure [43], photodoping [44–47], and some of them also induce superconductivity [24,41,43].

In the following sections, we elaborate on the electronic band structures in momentum space. To account for the in-plane anisotropy of the (3×1) ribbon-chains and (3×3) butterfly-like cluster phases, we use a special set of high-symmetry points ($\overline{\mathrm{M}}_1/\overline{\mathrm{M}}_2$ and $\overline{\mathrm{K}}_1/\overline{\mathrm{K}}_2$) and Brillouin zones (BZ) based on the primitive unit cell of trigonal 1*T* as depicted in Fig. 2(g).

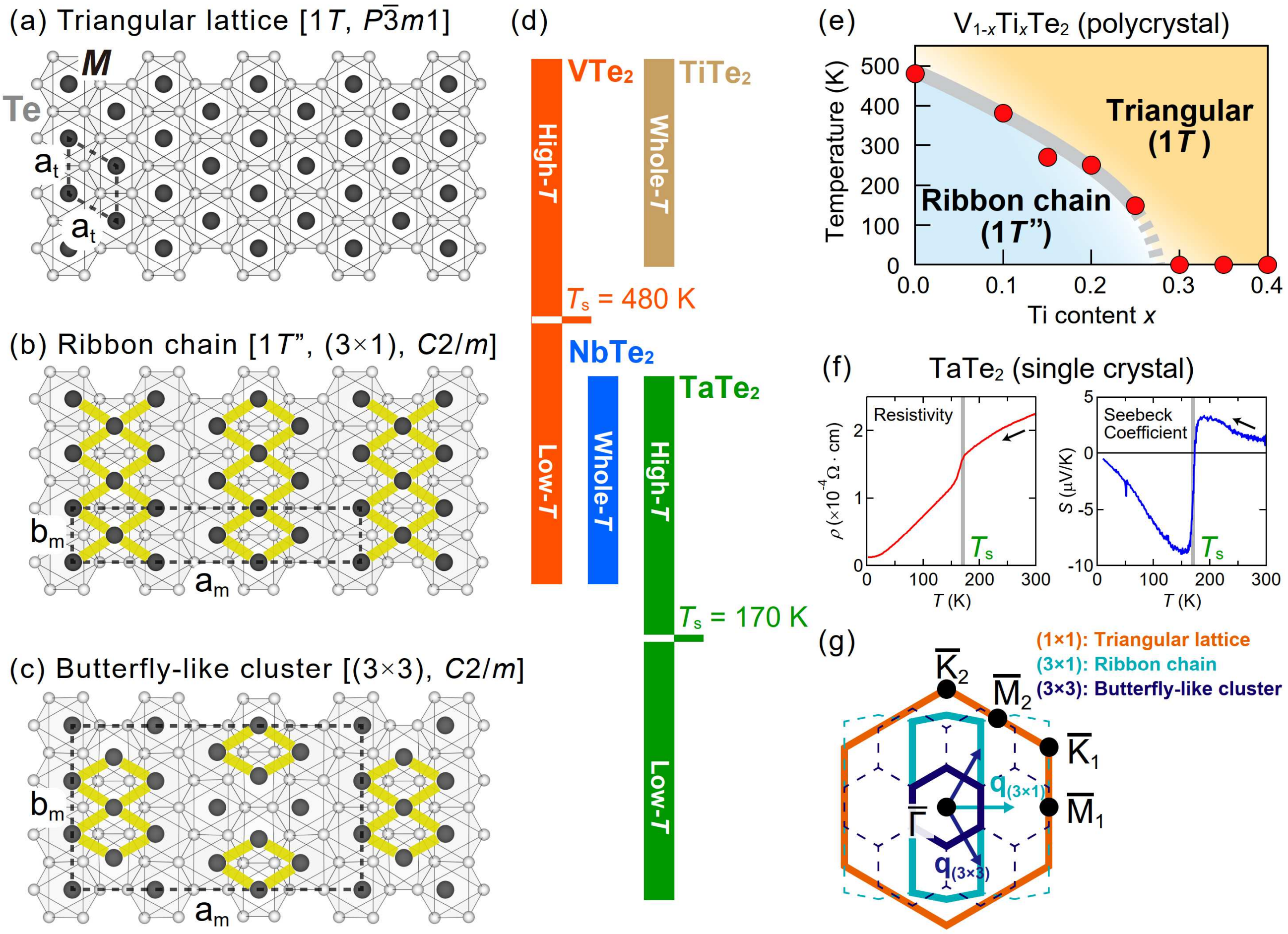

**Fig. 2. Phase transitions and superperiodic patterns in the group-5 *M*$Te_2$ (*M* = V, Nb, Ta).** (a)–(c) Top-down view of crystal structures for the undistorted trigonal 1*T* [(a), *P*3̄*m*1], ribbon chain (3×1) monoclinic 1*T*” [(b), *C*2/*m*], and butterfly-like cluster (3×3) monoclinic phases [(c), *C*2/*m*]. The broken lines represent the unit cells. The yellow lines in (b) and (c) show the characteristic ribbon chains and butterfly-like chains, respectively. (d) Thermal phase stabilities of *M*$Te_2$ (*M* = Ti, V, Nb, Ta). (e) Phase diagram of polycrystalline $V_{1-x}Ti_xTe_2$ derived from electric resistivity measurements [3]. (f) Temperature-dependent electric resistivity and Seebeck coefficient of single-crystalline $TaTe_2$. Adapted from Ref. [5] (© 2024 American Physical Society). (g) (001) surface Brillouin zones for the three phases shown in (a)–(c). $\mathbf{q}_{(3\times1)}$ and $\mathbf{q}_{(3\times3)}$ depict the reciprocal lattice vectors of the periodic lattice distortion for ribbon-chain and butterfly-cluster phases, respectively.

## 3. (V,Ti)$Te_2$: Trigonal 1*T* to Monoclinic Ribbon-chain 1*T*" Phase Transition

This section outlines the characteristics of the electronic band structures for the undistorted 1*T* and ribbon-chain 1*T*" phases based on the results in the pristine- and Ti-doped $VTe_2$. Section 3.1 presents the details of the Fermi surface and band structures in the 1*T* phase with special emphasis on their orbital characters. Section 3.2 highlights the anomalous flat bands and the resulting quasi-one-dimensional Fermi surface formed in the 1*T*" phase. Section 3.3 remarks the topological surface states realized in the (V,Ti)$Te_2$ system and their modifications through the 1*T*–1*T*" transition. Some parts of this section refer to our previous work in Ref. [26].

### 3.1 Fermi surfaces derived from $t_{2g}$ orbitals in the trigonal 1*T* phase

We begin with an overview of the electronic band structure in the undistorted trigonal 1*T*-(V,Ti)$Te_2$. Figure 3(a) represents the three-dimensional BZ in the 1*T* phase, where the blue arrows indicate the half-length reciprocal lattice vectors (**a***/2, **b***/2, **c***/2). Figure 3(b) shows the calculated electronic band structures of 1*T*-$VTe_2$ along the high-symmetry lines [see the red lines in Fig. 3(a)]. The size of the circular (diamond) markers depicts the contributions of V 3*d* (Te 5*p*) orbitals. The color of the circular markers represents the decomposition of the V 3*d* $t_{2g}$ orbital [$d_{XY}$/$d_{YZ}$/$d_{ZX}$, for the *XYZ* notation see Fig. 1(a)]. Along the BZ boundary (M–K/L–H lines), the linear band dispersion crossing the Fermi level ($E_F$) is found to be mainly derived from the V 3$d_{XY}$ orbital (green circles). Note that considering the trigonal symmetry, $d_{YZ}$ and $d_{ZX}$ orbitals are responsible for other two M–K/L–H lines. Meanwhile, we can see the Te 5*p* band forming a hole-like Fermi surface around Γ. These characteristic band structures in the calculation are clearly observed in the ARPES spectra of 1*T*-$V_{0.90}Ti_{0.10}Te_2$ (350 K, photon energy: 21.2 eV) as displayed in Fig. 3(c) (We note that the single-crystal $V_{0.90}Ti_{0.10}Te_2$ shows the 1*T*-1*T*" transition at a mild temperature of ~280 K [26]). The coexistence of the V 3*d* and Te 5*p* derived Fermi surfaces indicates the significant *p*–*d* charge transfer from Te to V, which increases the *d*-electron number from (nominal) 1 toward 4/3. We note that the chalcogen-derived hole band around Γ sinks below the Fermi level in the case of the group-5 sulfides/selenides such as 1*T*-$TaS_2$ [8,48] and 1*T*-$VSe_2$ [49].

We then discuss the three-dimensionality of the Fermi surface. Figure 3(d) displays the calculated Fermi surface for 1*T*-$V_{0.87}Ti_{0.13}Te_2$. This indicates that the Fermi surface has a finite dependence along **c*** axis (or $k_z$ dependence). To experimentally reveal this, we performed the photon energy ($h\nu$) dependent ARPES measurements on 1*T*-$V_{0.90}Ti_{0.10}Te_2$ (350 K) using synchrotron light sources. Figures 3(e) and 3(f) show the Fermi surface intensity maps at $k_z = 0$ and $\pi/c$ planes collected with $h\nu = 83$ and 102 eV photons, respectively. As guided by cyan curves, the Fermi surface topology critically depends on $k_z$: there are triangular hole-like Fermi surfaces centered at K in $k_z = 0$, whereas ellipses electron-like ones at L in $k_z = \pi/c$. Here we emphasize that the $E_F$-crossing V-shaped band exist along $\overline{\mathrm{K}} - \overline{\mathrm{M}} - \overline{\mathrm{K}}$ irrespective to the $k_z$ value. Figure 3(g) displays the cross-sectional Fermi surface plot along the $k_{//}$–$k_z$ plane that contains the M–L line [the yellow plane in Fig. 3(a)], obtained by scanning the incident photon energy ($h\nu$ = 48–90 eV). Here $k_{//}$ is the in-plane momentum direction parallel to the reciprocal lattice vector (**q**-vector) of the ribbon-chain 1*T*" (3 × 1) periodic lattice distortion (**a***/3). To test the Fermi surface nesting scenario, we extract the Fermi momentum [red markers in Figs. 3(e)–(g)] by Lorentzian fitting, and compare the observed Fermi surface nesting vectors ($\mathbf{q}_{\mathrm{FSN}}$, red arrows) with the **q**-vector of the 1*T*" (3×1×3) periodic lattice distortion ($|\mathbf{q}^{2\mathrm{D}}_{\mathrm{PLD}}| = |\mathbf{a}^*/3| \sim 0.67$ Å$^{-1}$, $|\mathbf{q}^{3\mathrm{D}}_{\mathrm{PLD}}| = |\mathbf{a}^*/3 + \mathbf{c}^*/3| \sim 0.74$ Å$^{-1}$, gray arrows) as summarized in Fig. 3(h). Given that $\mathbf{q}_{\mathrm{FSN}}$ is set parallel to $\mathbf{q}^{2\mathrm{D}}_{\mathrm{PLD}}$ [for Figs. 3(e) and 3(f)] or $\mathbf{q}^{3\mathrm{D}}_{\mathrm{PLD}}$ [for Fig. 3(g)], the size of the nesting vector ($|\mathbf{q}_{\mathrm{FSN}}|$) is evaluated to be at most $0.64(3)|\mathbf{q}^{2\mathrm{D}}_{\mathrm{PLD}}|$ for Fig. 3(e), $0.72(4)|\mathbf{q}^{2\mathrm{D}}_{\mathrm{PLD}}|$ for Fig. 3(f), and $0.80(3)|\mathbf{q}^{3\mathrm{D}}_{\mathrm{PLD}}|$ for Fig. 3(g). For all cases, $|\mathbf{q}_{\mathrm{FSN}}|$ is sufficiently smaller compared to $|\mathbf{q}^{2\mathrm{D}}_{\mathrm{PLD}}|$ or $|\mathbf{q}^{3\mathrm{D}}_{\mathrm{PLD}}|$, thus the simple Fermi surface nesting is not the origin of 1*T*-1*T*" phase transition.

Before introducing Whangbo's chemical bonding scenario, it is very instructive to classify the obtained Fermi surfaces near the BZ boundaries based on their orbital characters. At sufficiently close to the BZ boundaries, the $k_z$-dependence of the Fermi surfaces is quite small, and they can be simplified as a superposition of the three *virtual* one-dimensional Fermi surfaces as illustrated in the top part of Fig. 3(i). Given that the 1*T* structure has a threefold trigonal symmetry, each of the *virtual* one-dimensional Fermi surfaces are derived from one of the $t_{2g}$ ($d_{XY}$, $d_{YZ}$, or $d_{ZX}$) σ-bonding state spanning along the edge-sharing octahedral network [see the bottom part of Fig. 3(i)]. The resulting *virtual* one-dimensional band dispersion should exhibit a dispersive V-shape along $\overline{\mathrm{K}} - \overline{\mathrm{M}} - \overline{\mathrm{K}}$, which is parallel to the σ bonding. This simple concept represents the *hidden* Fermi surface raised by Whangbo *et al*. [10,50] as mentioned in Sect.1, which makes a good correspondence with the first principles calculation [Fig. 3(b)] as long as focusing on the BZ boundary. It has been also widely adopted in inorganic compounds including 1*T*-$MX_2$ [14,26,51] and purple bronzes $A\mathrm{Mo}_6\mathrm{O}_{17}$ ($A$ = Na, K) [50,52,53].

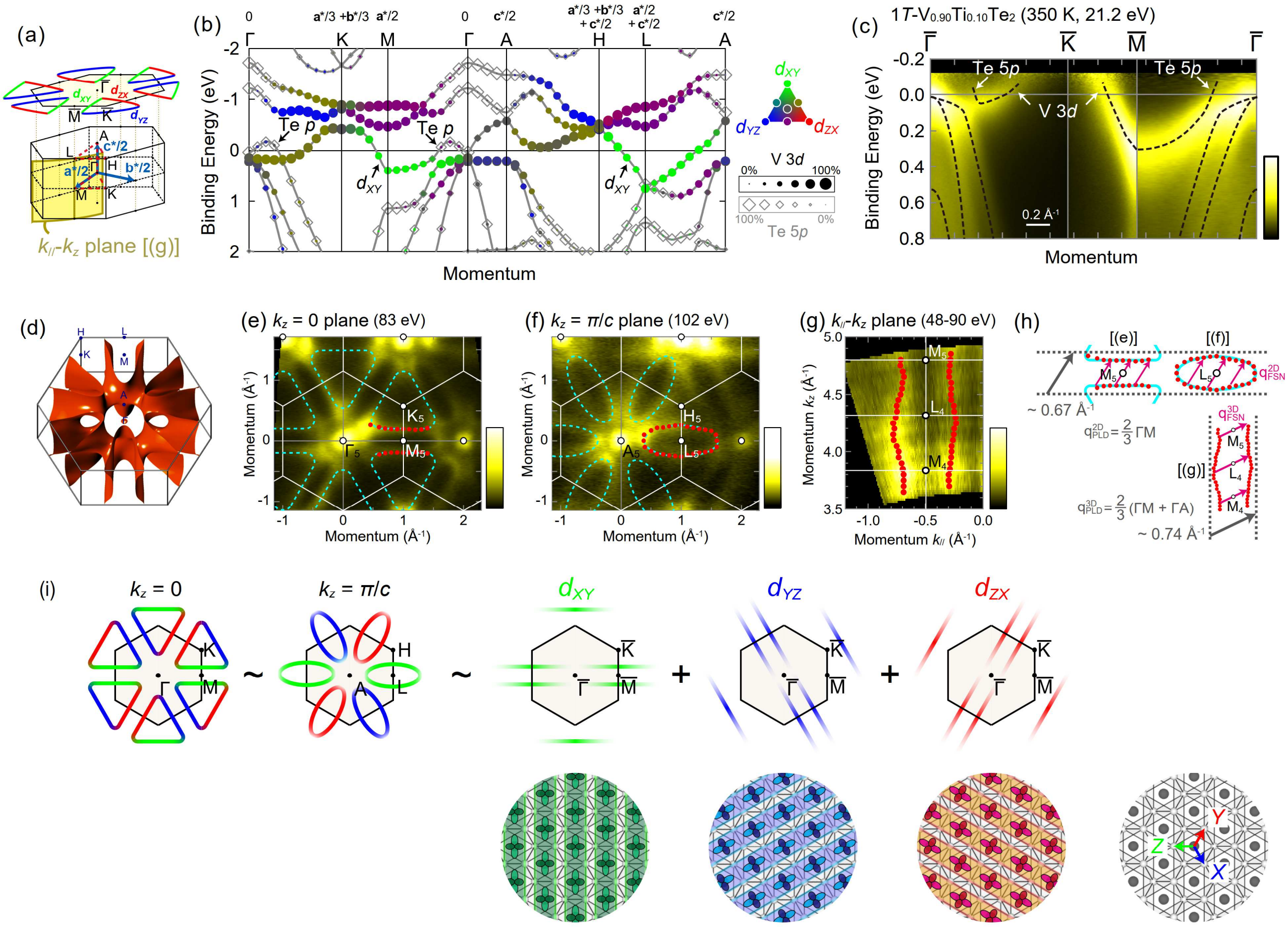

**Fig. 3. Band structure and Fermi surface in the trigonal 1*T*-(V,Ti)$Te_2$.**

(a) Brillouin zone for the trigonal 1*T*. (b) Orbital-weighted band calculations along high-symmetry lines for 1*T*-$VTe_2$. Circular (diamond) markers indicate the contribution from the V 3*d* (Te 5*p*) orbitals. (c) APRES data along the $\bar{\Gamma}-\bar{K}-\bar{M}-\bar{\Gamma}$ direction collected in 1*T*-$V_{0.90}Ti_{0.10}Te_2$ (350 K, 21.2 eV). (d) Calculated Fermi surface contours for 1*T*-$V_{0.87}Ti_{0.13}Te_2$. (e), (f) Fermi surface maps at $k_z = 0$ [(e)] and $\pi/c$ planes [(f)]. These data are recorded with synchrotron light sources (83, 102 eV). The red markers depict the Fermi momentum determined by the numerical fitting of their momentum distribution curves. The cyan broken lines trace the Fermi surface contours around the Brillouin zone boundaries. (g) Fermi surface map along the $k_x$-$k_z$ plane passing through M-L lines [see the yellow plane in (a)] recorded using 48–90 eV photons (interval of 3 eV). (h) Comparison between the Fermi surfaces nesting vectors ($\mathbf{q}_{FSN}$) obtained from (d)–(f) and the **q**-vector of the 1*T*” (3×1×3) periodic lattice distortion ($\mathbf{q}_{PLD}$), (i) Illustration of the one-dimensional *hidden* Fermi surfaces derived from the orthogonal V $t_{2g}$ σ-bonding. (a)–(d) are adopted and edited from Ref. [3] (© 2020 Springer Nature).

### 3.2 Flat band and quasi-one-dimensional Fermi surface induced by the $d_{YZ}/d_{ZX}$ trimerization the 1*T*” phase

Next, we highlight the anisotropic electronic modification emerging across the ribbon-chain formation observed in the (V,Ti)$Te_2$ system. Figures 4(a), (b) and 4(c), (d) display the ARPES results of the Fermi surface and band dispersions recorded with synchrotron light sources for 1*T*-$V_{0.90}Ti_{0.10}Te_2$ (350 K, $h\nu$ = 83 eV) and 1*T*”-$VTe_2$ (20 K, $h\nu$ = 86 eV), respectively. As described in Sect. 3.1, the Fermi surfaces of 1*T* at $k_z = 0$ [Fig. 4(a)] are characterized by the Te 5*p*-derived hole pocket at $\bar{\Gamma}$ and V 3*d*-derived large triangular ones centered at $\bar{\mathrm{K}}$. The latter originates from the V-shaped band dispersions spanning along $\bar{\mathrm{K}}-\bar{\mathrm{M}}-\bar{\mathrm{K}}$ [Fig. 4(b)]. Across the 1*T* to 1*T*” phase transition, the rotational symmetry changes from threefold ($P\bar{3}m1$) to twofold ($C2/m$) and accordingly the originally equivalent three $\bar{\mathrm{M}}$ points become one $\bar{\mathrm{M}}_1$ and two $\bar{\mathrm{M}}_2$ points. As shown in Fig. 4(d), the original $E_F$-crossing V-shaped band dispersions along $\bar{\mathrm{K}}-\bar{\mathrm{M}}-\bar{\mathrm{K}}$ are transformed into the flat bands along $\bar{\mathrm{K}}_1-\bar{\mathrm{M}}_2-\bar{\mathrm{K}}_2$. On the other hand, the V-shaped band is retained along $\bar{\mathrm{K}}_1-\bar{\mathrm{M}}_1-\bar{\mathrm{K}}_1$, resulting in a strongly anisotropic quasi-one-dimensional Fermi surface contour as demonstrated in Fig. 4(c). Thus, the 1*T*-1*T*” phase transition accompanies the huge directional modification of the electronic structure.

These anisotropic band modifications can be essentially understood by the trimerization picture of specific V 3*d* $t_{2g}$ orbitals. Here we set the *Z* axis as the unique V-Te bond direction which is perpendicular to the vanadium ribbon chain (i.e., $\mathbf{b}_m$ axis) as shown in Fig. 4(e). Figure 4(f) displays the calculated partial density of states (PDOS) of the V $t_{2g}$ orbitals in 1*T*/1*T*”-$VTe_2$. In the trigonal 1*T* phase, the equivalent three $t_{2g}$ states are broadly distributed in the vicinity of the Fermi level, and each forms the $E_F$-crossing V-shaped band at different BZ boundaries. Through the 1*T*-1*T*” transition, the two of three $t_{2g}$ orbitals ($d_{YZ}$ and $d_{ZX}$) are split into three segments, lying around $E_B$ ~ 0.4, −0.5, and −1.3 eV. Recalling the “three-center two-electron” picture depicted in the right part of Fig. 1(b), they are reminiscent of bonding (“B”), non-bonding (“NB”), and anti-bonding (“AB”) states in molecular orbitals of linear trimer, arising from the local σ-bonding of $d_{YZ}/d_{ZX}$ orbitals in the adjacent three vanadium atoms [see the pink oval in Fig. 4(e) for the $d_{YZ}$ case]. Accordingly, the $d_{YZ}/d_{ZX}$-derived itinerant V-shaped bands transform into the localized flat ones around $\bar{\mathrm{M}}_2$ as the bonding states below the Fermi level. Meanwhile, the remaining $d_{XY}$ state stays relatively intact, and the corresponding V-shaped band at the $\bar{\mathrm{M}}_1$ side survives throughout the phase transition. Consequently, the orbital-dependent modification on the Fermi surface can be illustrated as Fig. 4(g).

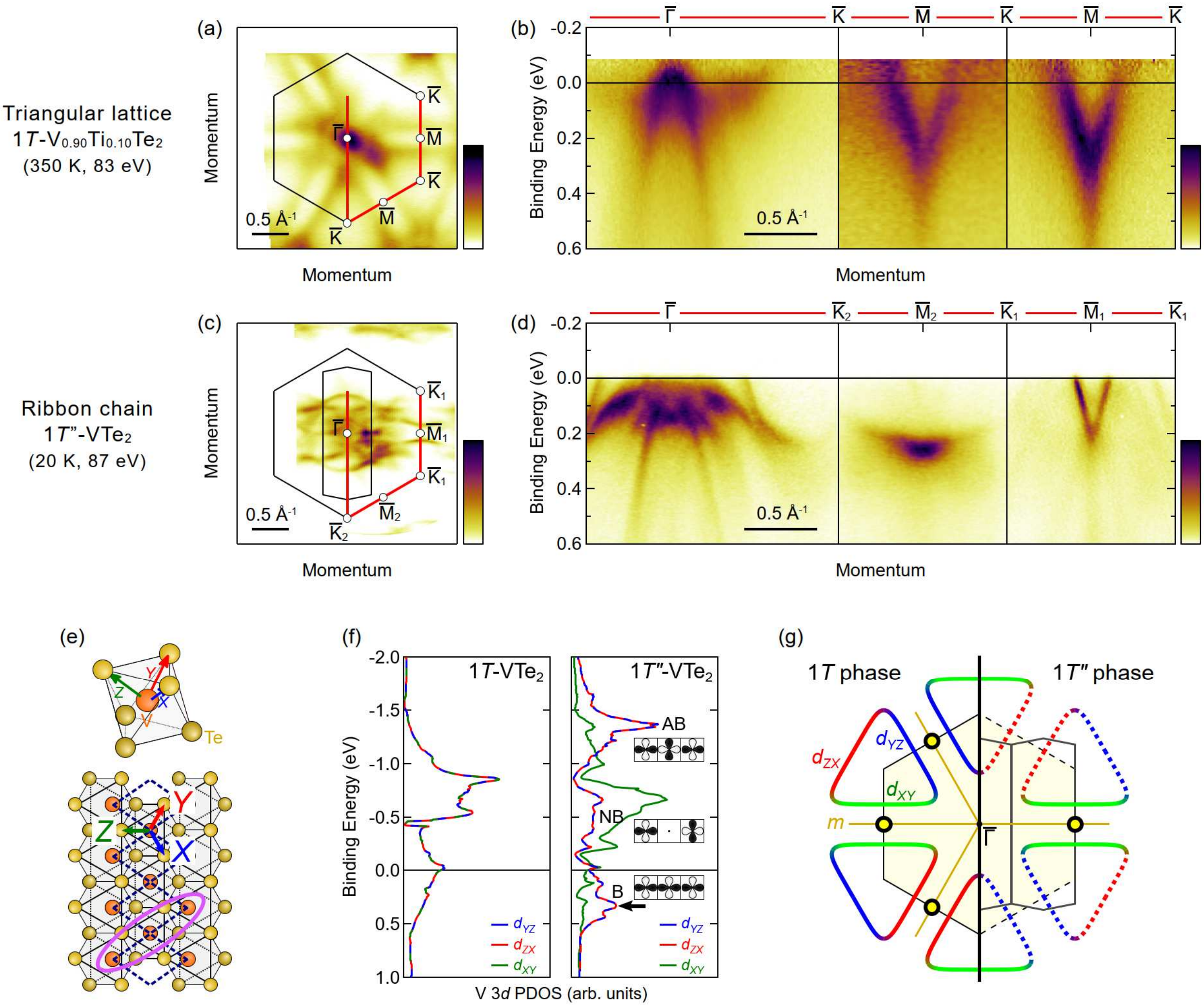

**Fig. 4. Formation of $d_{YZ}/d_{ZX}$ flat bands and quasi-one-dimensional Fermi surface observed in the ribbon chain 1*T*" phase (V,Ti)$Te_2$ system.**

(a) ARPES intensity map at $E_F$ in the triangular 1*T* phase of $V_{0.90}Ti_{0.10}Te_2$. (b) ARPES spectra along high-symmetry lines as depicted by the red lines in (a). Data are collected at 350 K (above the 1*T*-1*T*" transition temperature of ~280 K) using 87 eV photons. (c), (d) Same as (a), (b), but collected for the monoclinic 1*T*" phase of $VTe_2$ (20 K, 83 eV). (e) Local orthogonal *XYZ* coordination. Here we set *Z* as the direction perpendicular to the vanadium's ribbon chains. The pink oval indicates the trimerization of $d_{YZ}$ orbitals. (f) Calculated partial density of states (PDOS) for V 3*d* $t_{2g}$ orbitals in 1*T* /1*T*"-$VTe_2$. "B", "NB", "AB" represent the bonding, non-bonding, and anti-bonding states realized in 1*T*", respectively. The arrow indicates the $d_{YZ}/d_{ZX}$ PDOS peak assigned to the flat bands. (g) Cartoon of the Fermi surface reconstruction around the Brillouin zone boundaries. "*m*" indicates the mirror planes. (e)–(g) are adapted from Ref. [26] (© 2020 Springer Nature).

### 3.3 Selective disappearance of topological surface states through the 1*T*-1*T*" transition

Another intriguing aspect of the 1*T*-1*T*" transition in (V,Ti)$Te_2$ is its significant impact on the nontrivial topological surface states around the BZ boundary. Here, the key to the topological band inversion is the sufficient $k_z$-dependence of bulk bands along the M–L direction [see Fig. 5(a) for BZ]. Figure 5(b) displays the orbitally resolved band calculations for 1*T*-$V_{0.87}Ti_{0.13}Te_2$ at selected $k_z$ planes along the momentum parallel to Γ–M ($k_{\Gamma M}$). The colors of the curves indicate the weight of atomic orbitals as depicted by the color scale (note that the *xyz* coordinate used here is different from the octahedral *XYZ* axis). The black broken curves represent the results in the absence of spin-orbit interaction. Focusing on the topmost two bands at M and L, denoted as "A" and "B", their orbital characters of mainly V 3*d* (blue-like, even parity) and Te 5$p_x$+$p_y$ (red-like, odd parity) get inverted at a specific $k_z$ (~0.76 $\pi/c$). Since the $k_{\Gamma M}$ resides in a mirror plane, these two bands cannot hybridize or gap out without the help of spin-orbit interaction [see the broken curves in Fig. 5(b)]. Upon introduction of the spin-orbit interaction, the topological band inversion occurs and gives rise to the Dirac cone-like surface state within the bulk band gap (~0.2 eV) along $\overline{\mathrm{K}}-\overline{\mathrm{M}}-\overline{\mathrm{K}}$ on a (001) surface, as presented by the slab calculation shown in Fig. 5(c). Such a $k_z$ band inversion is indicative of the strong *d*–*p* hybridization characteristic to tellurides, which is seemingly absent in 1*T*-$VS_2$ [54] and 1*T*-$VSe_2$ [55] (although note that a recent study of strain-applied $VSe_2$ [56] predicts as similar band inversion mechanism).

Figures 5(d) and 5(e) summarize the ARPES results that highlight the Fermi surface and bulk/surface bands (purple/cyan curves) at the BZ boundaries recorded in the normal-state 1*T*-$V_{0.87}Ti_{0.13}Te_2$ (300 K, 21.2 eV) and the CDW-state 1*T*"-$VTe_2$ (200 K, 90 eV), respectively. In the 1*T* phase [Fig. 5(d)], the Dirac cone-like topological surface state together with the two bulk bands "A" and "B" are observed to be similar to those predicted by the slab calculation [Fig. 5(c)]. Thorough *hν*-dependent ARPES measurements [Fig. 5(f)] reveal the two-dimensional nature of the Dirac surface states and the substantial $k_z$ dispersion of the bulk bands "A" and "B". We note that the bulk band "A" is identical to the $t_{2g}$-dominated V-shaped band that forms the Fermi surface at the BZ boundary. In the 1*T*" phase [Fig. 5(e)], a similar Dirac surface state is preserved at the $\overline{\mathrm{M}}_1$ side, but disappears completely at the $\overline{\mathrm{M}}_2$ side. These contrasting features can be attributed to the $k_z$ dependence of the corresponding bulk bands as shown in Figs. 5(g) and 5(h). At the $\overline{\mathrm{M}}_1$ side [Fig. 5(g)], the V-shaped band clearly shows the finite $k_z$-dispersion together with the *hν*-independent Dirac surface state akin to those in the 1*T* phase [Fig. 5(f)]. On the other hand, the flat band at the $\overline{\mathrm{M}}_2$ side associated with trimerization exhibits negligible $k_z$-dependence [Fig. 5(h)], leading to the dissolution of the topological band inversion. This experimental finding demonstrates a unique way to manipulate topological states by molecular-like local bonding, though further theoretical analysis of the topological phase in 1$T''$-VTe$_2$ is desired.

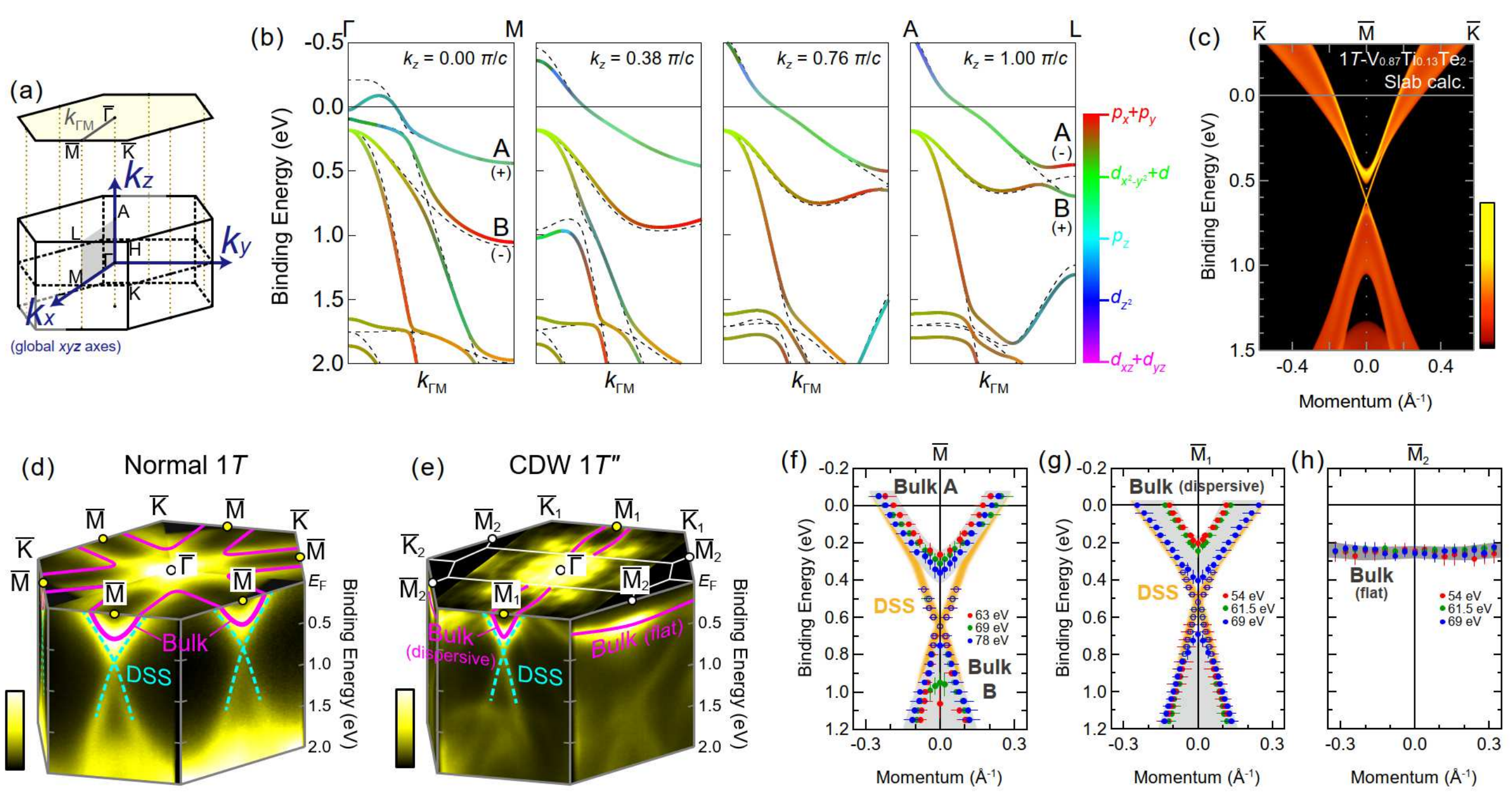

**Fig. 5. Significance of the bulk $k_z$ dispersion for the topological surface states in (V,Ti)Te$_2$.**

(a) Three-dimensional Brillouin zone of the trigonal 1$T$. ($k_x$, $k_y$, $k_z$) axes span the reciprocal space corresponding with the global *xyz* axes adopted for the calculation in (b). (b) Orbital-projected band calculation for 1$T$-V$_{0.87}$Ti$_{0.13}$Te$_2$. The broken curves indicate the results without spin-orbit coupling. (c) Corresponding (001) slab calculation along $\overline{\mathrm{K}} - \overline{\mathrm{M}} - \overline{\mathrm{K}}$. (d), (e) Observation of the Fermi surface and bulk/surface bands for 1$T$-V$_{0.87}$Ti$_{0.13}$Te$_2$ [(d), 300 K, 21.2 eV] and 1$T''$-VTe$_2$ [(e), 200 K, 90 eV]. The purple and cyan curves depict the bulk bands ("Bulk") and Dirac surface states ("DSS"), respectively. (f)–(h) Testing the $k_z$-dependence of band structures around $\overline{\mathrm{M}}$ for 1$T$-V$_{0.90}$Ti$_{0.10}$Te$_2$ [(f)] and $\overline{\mathrm{M}}_1$/$\overline{\mathrm{M}}_2$ for 1$T''$-VTe$_2$ [(g) and (h)] from *hν*-dependent ARPES. Adapted from Ref. [26] (© 2020 Springer Nature).

## 4. $VTe_2$, $NbTe_2$, and $TaTe_2$: Common Ribbon-chain 1*T*" Phase

As previously described in Sect. 2, $M$Te$_2$ ($M$ = V, Nb, Ta) commonly crystallize in the ribbon-chain 1*T*" superstructure at room temperature, while only $VTe_2$ undergoes the 1*T* structure above ~480 K. This section aims to discuss the origin of the *M*-dependence of the 1*T*" phase instability with respect to the 1*T* phase from the viewpoint of electronic band structures. Section 4.1 presents the results of the thermal analysis measurements. Section 4.2 compares the electronic band structures in 1*T*"-$M$Te$_2$ ($M$ = V, Nb, Ta) and remarks the energy gain resulting from the flat band formation reflecting the trimerization.

### 4.1 Phase stability at high-temperature regime

To confirm the 1*T*"-1*T* phase transition, we performed the thermogravimetric and differential thermal analysis (TG-DTA) on single-crystalline $M$Te$_2$ ($M$ = V, Nb, Ta) using a NETZSCH TG-DTA2500-IW thermal analyzer. Several pieces of samples (10–15 mg in total) and the equivalent weight of α-$Al_2O_3$ powders (for reference) were loaded into open alumina pans and then heated from room temperature at a rate of 10 K/min with flowing inert nitrogen gas (70 ml/min).

Figures 6(a)–(c) show the TG (weight loss) and DTA (heat flow) profiles for the three compounds. For $VTe_2$ [Fig. 6(a)], the DTA profile exhibits an endothermic peak at around 480 K without any weight change. This is a fingerprint of the first-order 1*T*-1*T*" phase transition as reported in the previous literature [29]. Upon further heating, a weight loss occurs at around 700 K, indicating that the 1*T*-$VTe_2$ crystals decompose above this temperature regime. Similar decomposition behavior is seen in $NbTe_2$ [Fig. 6(b)] and $TaTe_2$ [Fig. 6(c)] above around 800 and 750 K, respectively. On the other hand, their DTA profiles do not exhibit any endothermic peaks indicative of the 1*T*-1*T*" transition, thus the room-temperature 1*T*" configuration is retained up to the thermal decomposition temperature for $NbTe_2$ and $TaTe_2$.

### 4.2 Universal flat band formation at different energy levels

Figures 6(d)–(f) show the calculated band dispersions at $k_z$ = 0 for 1*T*"-$M$Te$_2$ ($M$ = V, Nb, Ta). The band structures in overall are very complicated for all compounds due to the tripling of the unit cell compared to the undistorted trigonal 1*T*. Nevertheless, they commonly host the dispersive-less flat band structures along $\bar{\mathrm{M}}_2 - \bar{\mathrm{K}}_2$ as marked by arrows. To experimentally characterize these flat bands, we collected the ARPES spectra along $\bar{\Gamma} - \bar{\mathrm{M}}_2 - \bar{\mathrm{K}}_2$ by using a He-discharge lamp ($h\nu$ = 21.2 eV) for $VTe_2$ [Fig. 6(g), 15 K], $NbTe_2$ [Fig. 6(h), 15 K], and $TaTe_2$ [Fig. 6(i), 300 K]. Here we note that the spectra of $NbTe_2$ [Fig. 6(h)] includes some minor photoelectron signals from the $\bar{\Gamma} - \bar{\mathrm{M}}_1 - \bar{\mathrm{K}}_1$line due

to the mixing of in-plane 120° domains [23] within the sample. For all three compounds, a pair of flat bands exist around the $\overline{\mathrm{M}}_2$ point. Thus the flat band formation is universal in 1*T*”-*M*Te$_2$. Meanwhile, their energy positions for VTe$_2$ (0.25/0.38 eV at $\overline{\mathrm{M}}_2$, see the energy distribution curve in Fig. 6(j)) are located at significant shallower $E_{\mathrm{B}}$ compared to those for NbTe$_2$ (0.59/0.88 eV) and TaTe$_2$ (0.62/0.91 eV). This trend is well reproduced by the calculations shown in Figs. 6(d)–(f). Although the corresponding energy positions in the calculations for NbTe$_2$ (0.63/0.85 eV) and TaTe$_2$ (0.64/0.84 eV) are close to those observed by ARPES, there is a substantial energy mismatch between the calculation (0.37/0.46 eV) and experiment (0.25/0.38 eV) for VTe$_2$.We ascribe this mismatch to the correlation effects of the V 3*d* electrons, which should be stronger compared to Nb 4*d* and Ta 5*d*. We note that the electron correlation effect had been also discussed in ARPES studies on other vanadium-based dichalcogenides 1*T*-VS$_2$ [57] and 1*T*-VSe$_2$ [49,58], where the narrowing of bandwidth were found in comparison with the band calculation.

As described for the (V,Ti)Te$_2$ case in Sect. 3.2, the 1*T*–1*T*” phase transition accompanying the trimerization of $d_{YZ}$/$d_{ZX}$ orbitals transforms the $E_{\mathrm{F}}$-crossing V-shaped bands into the flat bands along $\overline{\mathrm{K}}_1-\overline{\mathrm{M}}_2-\overline{\mathrm{K}}_2$ as illustrated in Fig. 7(a). Such a fully “gapped” band modification should result in lowering the energy of involved electrons as marked by arrows in Fig. 7(a). Here we discuss the energy of the flat bands based on the band calculations for VTe$_2$ and TaTe$_2$. Figures 7(b) and 7(c) display the PDOS distributions for $d_{YZ}$ [(b)] and the band dispersions along $\overline{\mathrm{K}}_1-\overline{\mathrm{M}}_2-\overline{\mathrm{K}}_2$ [(c)] in the 1*T*/1*T*” phases for VTe$_2$. As discussed in Sect. 3, the PDOS shows a strong modification into three segmentations reflecting the trimerization, whereas the total bandwidth is rather weakly modified (~10% increase). Accordingly, the energy position of the flat band at $\overline{\mathrm{M}}_2$ in 1*T*” is roughly pinned to the bottom of the original V-shaped band at $\overline{\mathrm{M}}$ in 1*T*, as shown in Fig. 7(c). This suggests that the corresponding eigenstates at $\overline{\mathrm{M}}_2$ and $\overline{\mathrm{M}}$ for 1*T*” and 1*T*, respectively represented by the trimerized and uniform σ-bonding $d_{YZ}$ chains, have similar eigenenergies. At the same time, the almost completely flat band realized for whole $\overline{\mathrm{M}}_2-\overline{\mathrm{K}}_2$ indicates the decoupled nature of the trimers, *i.e.* the negligible inter-trimer transfer integrals compared to intra-trimer ones. This trend is similarly found in the calculation of virtual 1*T*- and 1*T*”-TaTe$_2$ [Figs. 7(d) and 7(e)], but with the total bandwidth about 1.5 times wider than that of VTe$_2$. It reflects the itinerant nature of 5*d* orbitals compared to 3*d*. We should also note that the narrowing of the bandwidth caused by the electron correlation effects in VTe$_2$ further pushes the flat bands toward the Fermi level. The robustness of the 1*T*″ phase in (Nb,Ta)Te$_2$ as compared to VTe$_2$ is thus likely due to their energetically deeper flat bands [Fig. 6(j)] arising from the larger overlap of the *d* orbitals, even though one should consider the total energy balance between the gain of the electron system in the whole momentum space and the associated elastic cost in the lattice system.

Finally, we comment on the remarkable difference between $NbTe_2$ and $TaTe_2$. Although the present results in Fig. 6 appear quite similar, they rather exhibit distinct physical properties at low temperatures as described in Sect. 2. The above discussion is thus limited to the 1*T*” phase stability at high-temperature regime, with respect to the 1*T* phase. It is also noteworthy that they have different phase diagrams under high pressure. $NbTe_2$ undergoes a lattice-collapse phase transition accompanied by Nb dimerization at ~20 GPa [59]. In contrast, $TaTe_2$ does not show such structural phase transition while exhibiting three superconducting phases (up to ~3.5 K) upon increasing pressure [43]. Given that $NbTe_2$ has recently been proposed as a topological superconductor candidate [37,60], further high-resolution ARPES experiments using single-domain samples are highly desirable in order to clarify the relevant low-energy electronic band structures.

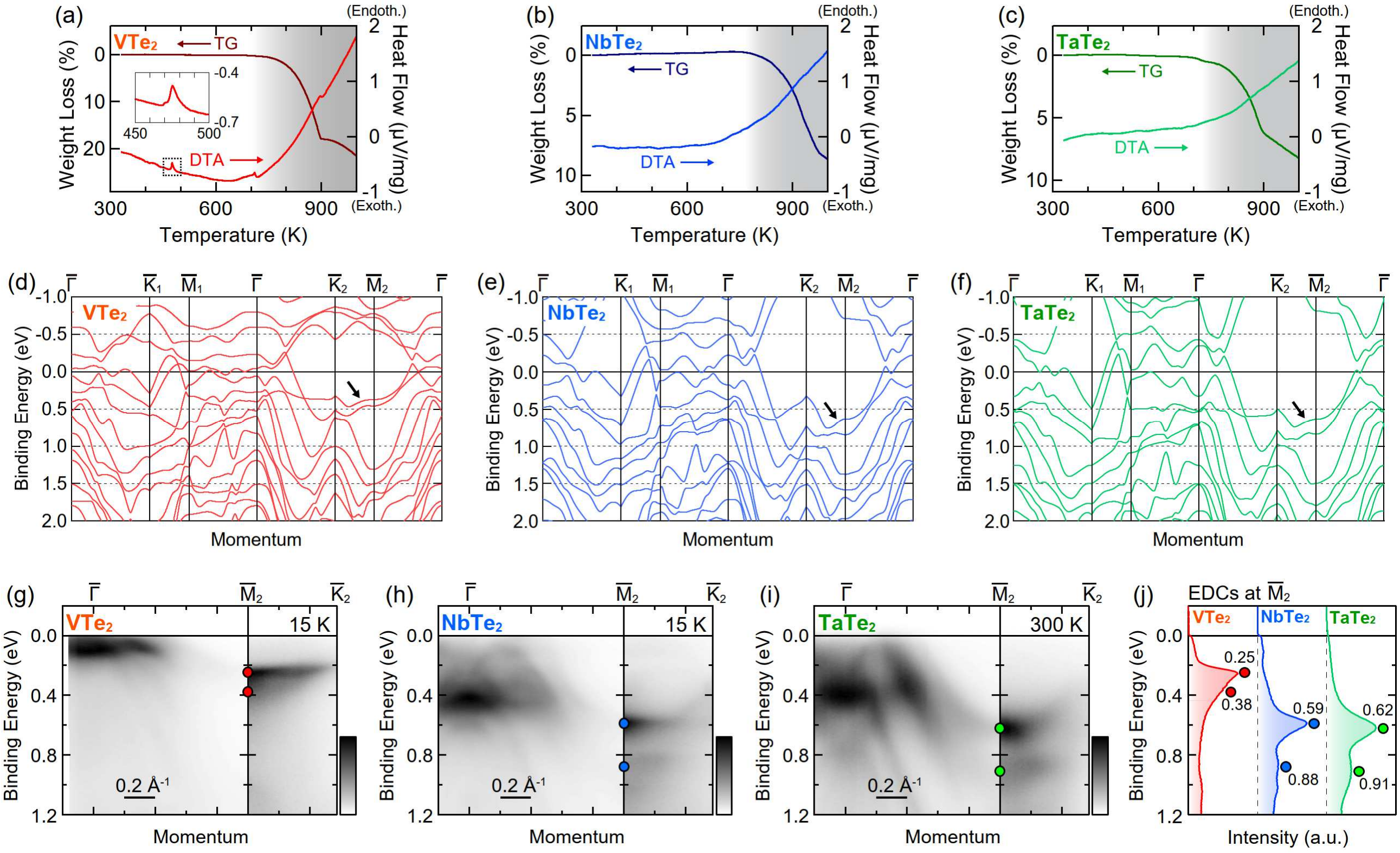

**Fig. 6. Universal flat band formation and relative phase stabilities in 1*T”*-*M*$Te_2$ (*M* = V, Nb, Ta).** (a)–(c) TG-DTA profiles for $VTe_2$ [(a)], $NbTe_2$ [(c)], and $TaTe_2$ [(c)]. The positive (negative) values of the heat flow indicate endothermic (exothermic) changes. The inset in (a) shows the endothermic peak indicative of the 1*T*–1*T”* transition. The gray shades highlight the decomposition temperature regime as found from the occurrence of weight loss. Data were recorded on heating from room temperature with a rate of 10 K/min under inert $N_2$ gas flow. (d)–(f) Calculated band dispersions for the 1*T”* phase in $VTe_2$ [(d)], $NbTe_2$ [(e)], and $TaTe_2$ [(f)]. The characteristic flat bands are depicted by arrows. (g)–(i) ARPES spectra along $\overline{\Gamma} - \overline{\mathrm{M}}_2 - \overline{\mathrm{K}}_2$ for $VTe_2$ [(g), 15 K], $NbTe_2$ [(h), 15 K], and $TaTe_2$ [(i), 300 K], acquired using a He-discharge lamp ($h\nu$ = 21.2 eV). The markers indicate the energy positions of flat bands. The spectra of $NbTe_2$ include minor photoelectron signals along $\overline{\Gamma} - \overline{\mathrm{M}}_1 - \overline{\mathrm{K}}_1$ due to the mixing of in-plane 120 deg. domains. (j) Comparison of energy distribution curves at $\overline{\mathrm{M}}_2$ (integral width: 0.05 Å$^{-1}$).

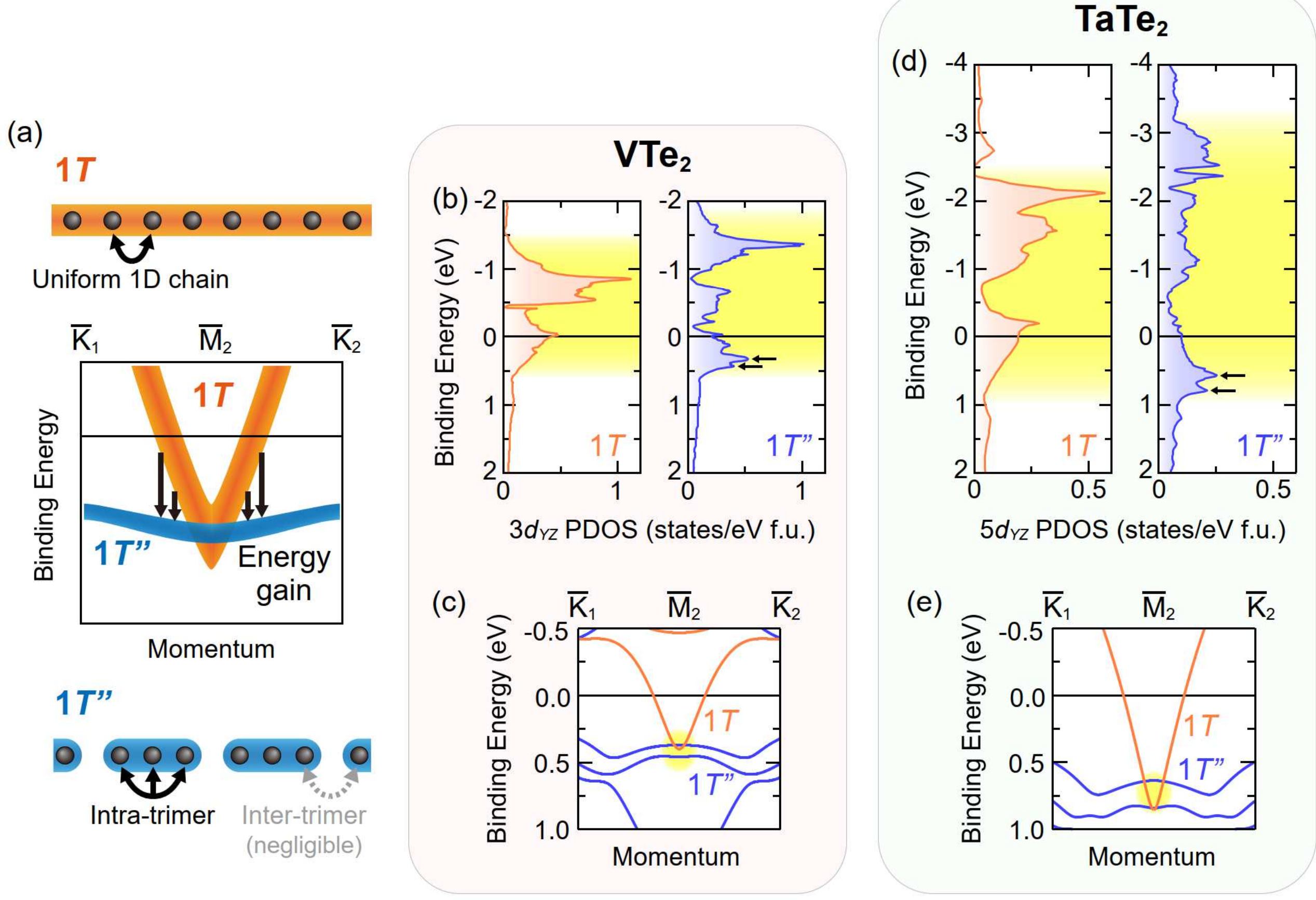

**Fig. 7. Energy gain stemming from the flat band formation in the 1*T*" phase.**

(a) Sketch of the one-dimensional *M* sites and the band dispersions along $\overline{\mathrm{K}}_1 - \overline{\mathrm{M}}_2 - \overline{\mathrm{K}}_2$ for the 1*T* and 1*T*" phases. The arrows in the band dispersions highlight the energy gain arising from the flat band formation through the 1*T*–1*T*" transition. (b) Partial density of states of the V 3$d_{YZ}$ orbital in $VTe_2$ for the 1*T*/1*T*" phases. The yellow shades roughly highlight the bandwidth. (c) Band dispersion along the Brillouin zone boundary at $k_z = 0$ in 1*T*/1*T*"-$VTe_2$. The yellow marker represents the bottom of the $d_{YZ}$-derived bands. (d), (e) Same as (b) and (c), but for the Ta 5$d_{YZ}$ orbital in $TaTe_2$. (b), (c) and (d), (e) are reproduced from Ref. [26] and [27], respectively.

## 5. $TaTe_2$: Ribbon-chain 1*T*” to Butterfly-like-cluster Phase Transition

This section presents a review of the electronic structure of $TaTe_2$ with a particular focus on its modifications through the ribbon-chain to butterfly-like-cluster phase transition at ~170 K, as revealed by temperature-dependent photoemission spectroscopy measurements with first-principles calculations [27]. Section 5.1 summarizes several experimental findings including core-level spectra splitting, chemical potential shift, band spectral segmentation, and kink-like band reconstruction. Section 5.2 is dedicated to the characteristic band structures along the BZ boundary and their orbital-dependent reconstructions through the transition. Section 5.3 highlights the structural and electronic fluctuations that appear at room temperature peculiar to $TaTe_2$.

### 5.1 Electronic states modifications: Overview

Figures 8(a), (b) and 8(c), (d) show the ARPES data of the Fermi surface and band dispersion across the transition temperature (300 and 15 K) collected with a He-discharge lamp (21.2 eV). At room-temperature phase [Fig. 8(a)], the quasi-one-dimensional wrapped Fermi surfaces extend along the $\overline{\Gamma}-\overline{\mathrm{M}}_1$ direction, similar to those observed in 1*T*”-$VTe_2$ [Fig. 3(c)]. In contrast, the spectral weight of the Fermi surfaces remains largely unchanged across the transition [Figs. 8(a) and 8(c)], which is in stark contrast to the 1*T*–1*T*” phase transition in (V,Ti)$Te_2$ [see Figs. 3(a) and 3(c)]. Meanwhile, we can find several types of modifications upon cooling in the core-level spectra and band dispersions as follows.

(i) Splitting of Ta 4*f* core-level spectra: Figures 8(e) and 8(f) respectively show the temperature-dependence of the Ta $4f_{7/2}$ and Te $4d_{5/2}$ core-level spectra collected on cooling with synchrotron light source ($h\nu$ = 90 eV). The Ta $4f_{7/2}$ core-level spectra [Fig. 8(e)] exhibit a remarkable splitting through the transition up to ~0.31 eV (see the broken curves for fitting peaks), indicating an enhancement in the variation of the electron densities among the inequivalent Ta sites (i.e., *d*–*d* charge transfer). We note that this splitting energy size is not comparable to that of the Star-of-David-type tantalum dichalcogenides $TaS_2$ (~1.2 eV) [61] or $TaSe_2$ (~0.9 eV) [62]. In contrast, the Te $4d_{5/2}$ spectral distribution [Fig. 8(f)] hardly changed over the measured temperature. This indicates that the charge distribution among the Te sites (*p*–*p* charge transfer) is negligible.

(ii) Chemical potential shift: Although the shape of the Te $4d_{5/2}$ core-level spectral profile remains mostly unchanged, it shifts to the lower $E_B$ regime (up to about 45 meV) upon cooling [see the overlaid profiles in Fig. 8(f)]. As summarized in Fig. 8(g), this temperature-dependent energy shift is similarly observed in the Ta 4*f* spectra (purple triangles and cyan rectangles) and the valence band dispersions (gray shades). This sizeable energy shift is therefore ascribed to the temperature-dependent chemical potential shift.

(iii) Spectral weight segmentation due to the band-folding effect: Around the BZ center (e.g., $\bar{\Gamma} - \bar{M}_2/\bar{\Gamma} - \bar{K}_2$), the hole-like bands split into sharp submanifolds with small energy gaps (typically 50–150 meV), as indicated by the white arrows in Fig. 8(d). These characteristics are well reproduced by band-unfolding spectral calculations as shown in Fig. 8(h). Here, $\bar{\Gamma}'$ denotes the newly introduced BZ center stemming from the in-plane (3×3) superstructure of the low-temperature butterfly-like cluster phase. Because several replica bands, such as the downward convex dispersions depicted by the upper two arrows in Fig. 8(h), are periodically distributed at both the $\bar{\Gamma}$ and $\bar{\Gamma}'$ points, these modifications are essentially attributed to the band-folding effect.

(iv) Kink-like band reconstruction: In the $\bar{M}_1 - \bar{K}_1$ direction (BZ boundary), the $E_F$-crossing liner band [depicted by the yellow arrows in Figs. 8(b) and 8(d)] undergoes a significant transformation into the less-dispersive kink-like structure, exhibiting a 70% reduction in gradient. The kink-like structure exhibits a characteristic peak at $E_B$ ~ 70 meV as evaluated from its energy distribution curve. This energy scale is consistent with the temperature-dependent optical conductivity spectra, where a sharp peak at 800 $cm^{-1}$ (~100 meV) indicative of an interband transition emerges below the transition temperature [45]. This type of the band reconstruction cannot be explained by the simple band-folding effect. The origin of this band will be elucidated in Sect. 5.2.

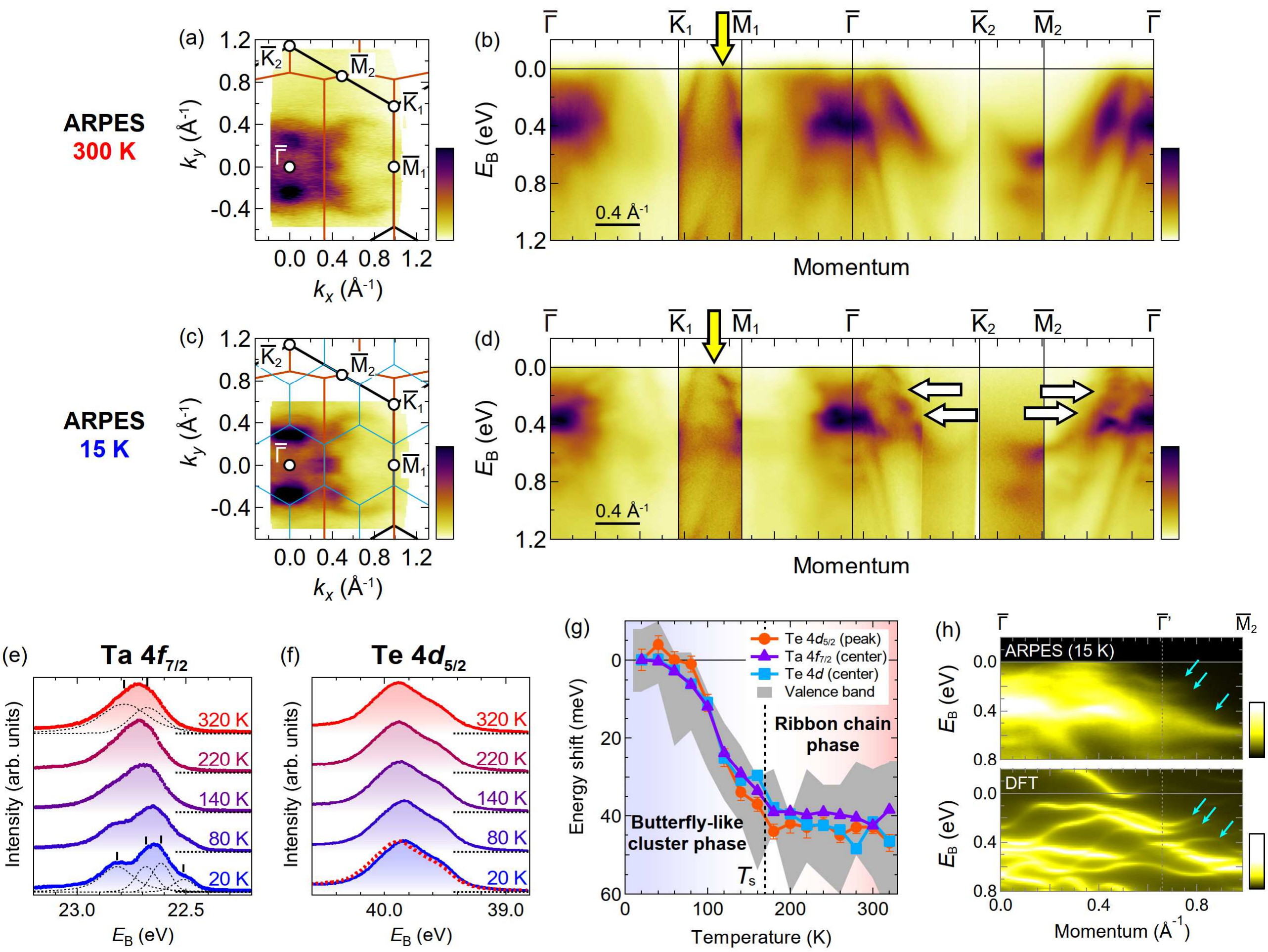

**Fig. 8. Electronic state modifications through the ribbon-chain to butterfly-like-cluster phase transition in $TaTe_2$.**

(a) ARPES intensity plots at $E_F$ (integral width: 20 meV) collected at 300 K ($hv$ = 21.2 eV). (b) ARPES spectra along $\overline{\Gamma}-\overline{\mathrm{K}}_{1(2)}-\overline{\mathrm{M}}_{1(2)}-\overline{\Gamma}$. (c), (d) Same as (a) and (b), but collected at 15 K. The white arrows mark the characteristic spectral segmentation features, whereas the yellow arrow depicts the kink-like band reconstruction appearing around $\overline{\mathrm{M}}_1$. (e), (f) Temperature evolution of the Ta 4$f_{7/2}$ [(e)] and Te 4$d_{5/2}$ [(f)] core-level spectra recorded in cooling scan using synchrotron light sources ($hv$ = 90 eV). The black dotted curves and markers in (e) represent the fitting Voigt functions and their peak positions, respectively. (g) Temperature dependence of energy position shifts (relative to the values at 20 K) for the highest intensity peak of Te 4$d_{5/2}$ (orange circles), the center of spectral weight of Ta 4$f_{7/2}$ (purple triangles) and Te 4$d$ (including 4$d_{5/2}$ and 4$d_{3/2}$, cyan rectangles). The gray shade shows the possible energy shift evaluated from the flat bands around $\overline{\mathrm{M}}_2$. (h) Close-up comparison of the ARPES spectra (15 K, 21.2 eV) and calculated band-unfolding spectra (at $k_z$ = 0) in the butterfly-like cluster phase. The cyan arrows mark characteristic sharp submanifold structures. $\overline{\Gamma}'$ denotes the newly introduced Brillouin zone center by the (3×3) superstructure. The cyan arrows depict the characteristic sharp submanifold structures. Adopted and edited from Ref. [27] (© 2024 American Physical Society).

### 5.2 Kink-like reconstruction in $d_{XY}$-derived band at Brillouin zone boundary

We now turn our attention to the orbital-dependent band reconstruction occurring along the BZ boundary. Figures 9(a)–(c) display the ARPES data (300 K, 21.2 eV), calculated band-unfolding spectra ($k_z$ = 0), and Ta 5$d$ $t_{2g}$ orbital-projected band calculation [with a same manner as Fig. 3(b)] along $\overline{\mathrm{K}}_1 - \overline{\mathrm{M}}_1 - \overline{\mathrm{K}}_1$ and $\overline{\mathrm{K}}_1 - \overline{\mathrm{M}}_2 - \overline{\mathrm{K}}_2$ in the ribbon-chain phase at room temperature. The contrasting band structures, namely the $E_F$-crossing V-shaped band at $\overline{\mathrm{M}}_1$ indicated by the green arrow and the two flat bands at $\overline{\mathrm{M}}_2$ by the blue and red arrows, are observed to be similar to those of 1$T$”-$VTe_2$ [Fig. 4(d)]. These bands are in good agreement with the band calculations [Figs. 9(b) and 9(c)] and originate from the Ta $d_{XY}$ and $d_{YZ}/d_{ZX}$ orbitals, respectively (see the green and red/blue arrows). Figures 9(d)–(f) show the results in the low-temperature butterfly-like cluster phase (ARPES data are collected at 15 K). The calculated band dispersion, represented by the thin lines in Fig. 9(e) and 9(f), show numerous branches due to the (3×3) folding. However, its unfolded band dispersion [Fig. 9(e)] exhibit similarities to the ARPES image [Fig. 9(d)]. Both datasets indicate that the $d_{YZ}/d_{ZX}$-dominated flat bands at $\overline{\mathrm{M}}_2$ are scarcely modified across the transition, whereas the $d_{XY}$-derived V-shaped band at $\overline{\mathrm{M}}_1$ undergoes a remarkable reconstruction into the unusual kink-like structure. Such $d$ orbital-dependent band reconstructions can also be seen as characteristic peaks in the Ta site-averaged PDOS distributions shown in Fig. 9(g), as indicated by the yellow arrows. The Ta site-specific PDOS calculations [27] further reveal that the middle string of Ta ribbon chains is primarily responsible for the kink-like $d_{XY}$ band reconstruction through the transition. Conversely, the Te states appear to remain largely unchanged throughout the transition, which is anchored by the temperature-independence of the Te 5$p$-derived band [depicted by the gray arrows in Figs. 9(a)–(f)] as well as the Te 4$d$ core-level spectra [Fig. 8(f)].

These results demonstrate the $d$ orbital-dependent electronic modifications through the ribbon-chain to butterfly-like-cluster phase transition in $TaTe_2$. However, the kink-like band reconstruction is arguably a milder change compared to the flat band formation around $\overline{\mathrm{M}}_2$ through the trigonal to ribbon-chain (1$T$-1$T$”) phase transition as observed in $(V,Ti)Te_2$. This may be explained by the fact that in the 1$T$” phase all $M$ atoms participate in the $d_{YZ}/d_{ZX}$ trimerization, whereas in the butterfly-like cluster phase only a subset of Ta atoms in the cluster is involved in the $d_{XY}$ kink structure (note that two-ninth of Ta atoms are isolated sites that are not included in clusters). Further studies are required to better understand the driving forces of the phase transition in $TaTe_2$ from both theoretical and experimental sides.

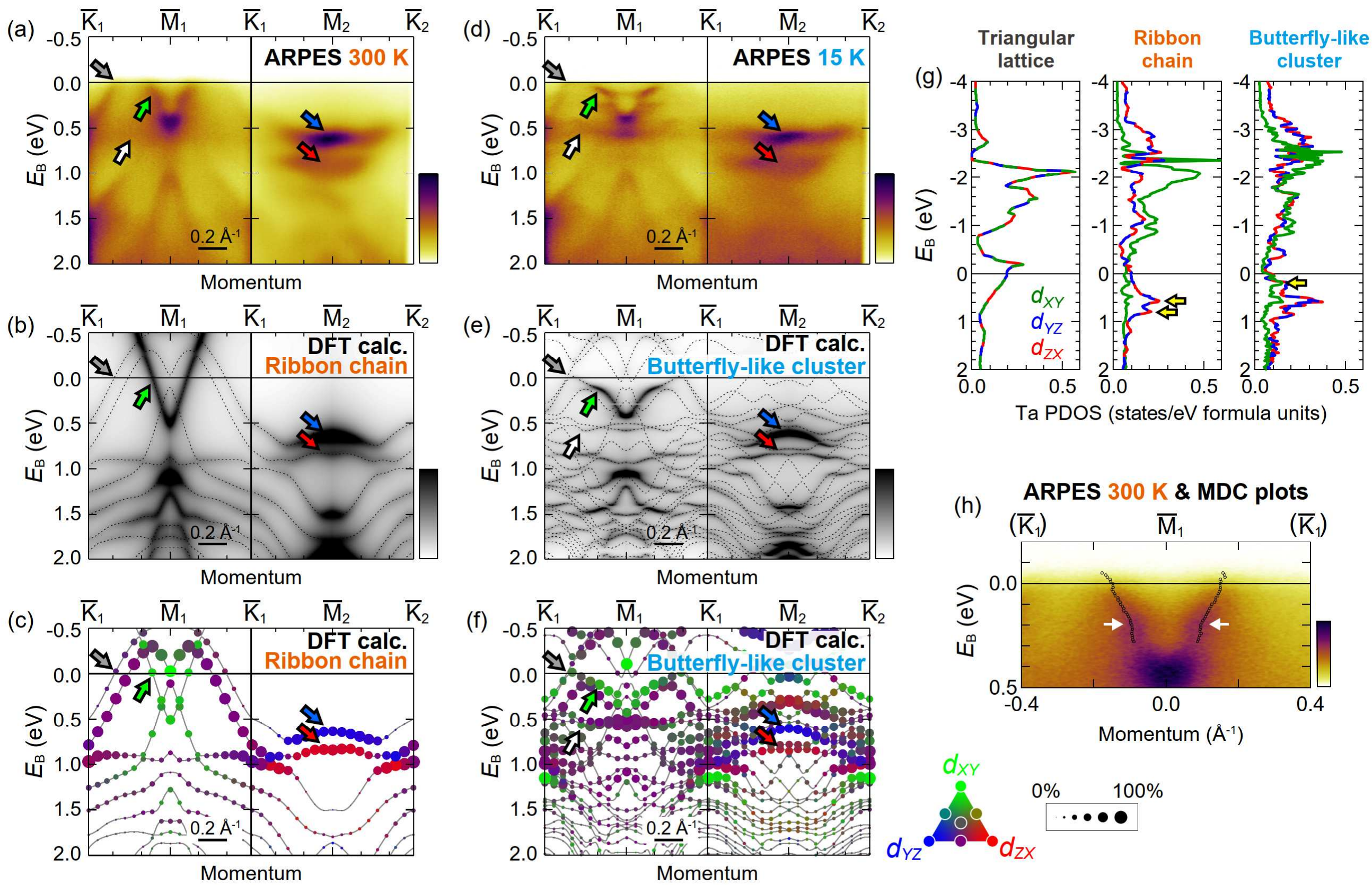

**Fig. 9. Band reconstructions along the Brillouin zone boundaries in $TaTe_2$.**

(a)–(c) ARPES data [300 K, (a)], band calculation with its unfolded spectra [(b)], and Ta $t_{2g}$ orbital-projected calculation [(c)] along $\overline{\mathrm{K}}_1 - \overline{\mathrm{M}}_1 - \overline{\mathrm{K}}_1$ and $\overline{\mathrm{K}}_1 - \overline{\mathrm{M}}_2 - \overline{\mathrm{K}}_2$ for the ribbon-chain phase. (d)–(f) Same as (a)–(c), but for the butterfly-like cluster phase (ARPES data: 15 K). The green and red/blue arrows mark the $d_{XY}$-dominated V-shaped band and the $d_{YZ/ZX}$-dominated flat bands, respectively, whereas the gray arrow indicates the Te $p$-derived $E_F$-crossing band. The while arrow depicts the band dispersion peculiar to the low-temperature phase but its remnant is found in the ARPES spectra at 300 K [(a)]. (g) Calculated PDOS of the Ta site-averaged $d_{XY}/d_{YZ}/d_{ZX}$ for the virtual trigonal 1$T$, room-temperature ribbon-chain 1$T$", and low-temperature butterfly-like cluster phases. The arrows in the middle panel mark the $d_{YZ}/d_{ZX}$ PDOS peaks in 1$T$" corresponding to the flat bands around $\overline{\mathrm{M}}_2$, whereas the one in the right panel kink-like band emerged at $\overline{\mathrm{M}}_1$ side. (h) Enlarged ARPES spectra at 300 K along $\overline{\mathrm{K}}_1 - \overline{\mathrm{M}}_1 - \overline{\mathrm{K}}_1$. The markers indicate the peak positions obtained by fitting the momentum distribution curves, tracking the faint kink-like feature (see also the white arrows). Adopted and edited from Ref. [27] (© 2024 American Physical Society).

### 5.3 Structural and electronic fluctuations emerging at room-temperature phase

While the ARPES results in $TaTe_2$ are basically in agreement with first-principles calculations, there are some signatures of structural/electronic fluctuations in the room-temperature ARPES data. For example, the spectral weight located at $E_B$ ~ 0.5 eV along $\overline{K}_1 - \overline{M}_1 - \overline{K}_1$ is unusually blurred, as indicated by the white arrow in Fig. 9(a). Furthermore, the $d_{XY}$-derived V-shaped band exhibits a faint kink-like feature even at 300 K, as evidenced by the peak plots in Fig. 9(h). Although anomalously broad, these features are characteristic of the butterfly-like cluster phase at low temperature [Figs. 9(e) and 9(f)] and are completely absent in the ribbon-chain phase calculations [Figs. 9(b) and 9(c)] under static atomic configurations.

From the perspective of structural analysis, the seminal work by Sörgel *et al.* [25] revealed that the room-temperature $TaTe_2$ exhibits an unusual large atomic displacement parameter ($U_{22}$ ~ 0.04 Å$^2$) in the middle string of Ta ribbon chains along the $\mathbf{b}_m$-axis. The recent comprehensive x-ray diffraction study on the group-5 $MTe_2$ ($M$ = V, Nb, Ta) by Katayama *et al.* [38] further reported that $TaTe_2$ hosts a conspicuous anisotropic structural fluctuation as compared to $VTe_2$ and $NbTe_2$ at room temperature. They also proposed that Ta atoms are locally displaced to dimerization with thermal fluctuations, instead of the conventional trimerization. In this context, the trimerization picture in the room-temperature $TaTe_2$ may not be as precise as that in $VTe_2$ or $NbTe_2$. Furthermore, it is noteworthy that an unusual convex-upward electrical resistivity curve [Fig. 2(f)] [25] and a broad Drude component in the optical conductivity spectrum [45] are observed above the transition temperature of 170 K, thereby suggesting that the electrons are strongly scattered in this regime. This indicates that the strong fluctuation in the middle string of Ta chains indeed affects the electronic band structure, leading to the anomalous electronic properties.

## 6. Summary and Outlook

In this review, we provided an overview of electronic structures realized in the group-5 ditellurides *M*$Te_2$ (*M* = V, Nb, Ta), based on the ARPES data and first-principles calculations. We highlighted the role of the localized molecular-like orbital bonds that form the unusual flat bands in specific momentum space, and discussed their impacts on the Fermi surface anisotropy, crystal phase stabilities, nontrivial topological properties, and so on. We believe that this review offers the comprehensive insight into the quantum electronic states in *M*$Te_2$ (*M* = V, Nb, Ta) and stimulates further studies on the electron-lattice-bond coupled phenomena emerging in various metallic systems.

Lastly, we would like to remark several potential avenues for future research on the group-5 *M*$Te_2$: (i) Recent studies revealed that the superstructures and electronic properties of *M*$Te_2$ are sometimes strongly affected by the slight change of non-stoichiometry, such as Te deficiency and/or *M* self-intercalations. Some of them include the different polymorphic (3×3) superstructure with antiferromagnetic ordering [63] and the Kondo effect [64] appearing in the vicinity of $VTe_2$. Fine tuning of such non-stoichiometry and possible chemical substitutions may lead to the new phases with novel polymorphism/polytypism in *M*$Te_2$ [41]. (ii) Electron-lattice-bond coupled phase in *M*$Te_2$ can be manipulated by external triggers, such as electric fields or optical pulses. Indeed, many works have been recently made in $VTe_2$ [30,32–35] and $TaTe_2$ [44–47]. These attempts will also provide promising routes to explore the nonequilibrium state of materials and realize exotic phenomena. (iii) The group-5 *M*$Te_2$ has great potential as a novel two-dimensional material. This is naively because reducing the thickness down to a monolayer should eliminate the moderately strong interlayer interaction inherent to tellurides as compared to sulfides and selenides, and affect the stability of trimerizations. Indeed, the bottom-up synthetic approaches such as molecular beam epitaxy (MBE) and chemical vapor deposition (CVD) have successfully fabricated mono/few-layer *M*$Te_2$ with different charge/spin orderings from the bulk counterparts (see Refs. [65–76] for $VTe_2$, [77,78] for $NbTe_2$, and [79–82] for $TaTe_2$). On the other hand, we would also encourage more research using the top-down exfoliation method to assemble van der Waals heterostructures and pursue the relevant novel phenomena such as twistronics. Although exfoliation down to the monolayer limit is challenging due to the strong interlayer coupling as compared to graphene, we expect that the state of the art exfoliation techniques such as the $Al_2O_3$-assisted mechanical exfoliation [83] and solid lithiation and exfoliation [84] will solve this issue.

**Acknowledgment**

The authors are grateful to Yusuke Sugita and Yukitoshi Motome (for providing band calculations of $NbTe_2$), Manabu Kamitani, Hidefumi Takahashi, Hideaki Sakai, and Shintaro Ishiwata (providing single-crystalline $NbTe_2$), Kiyoshi Nikaido and Tatsuo Hasegawa (assistance of TG-DTA measurements), Miho Kitamura, Koji Horiba, Hiroshi Kumigashira, and Kenichi Ozawa (set up of Photon Factory KEK BL28A), and Naoyuki Katayama (discussion about crystal structures). The synchrotron ARPES experiments were conducted under KEK-PF proposals (No. 2018G624 and 2023G108). This work was partly supported by Grands-in-Aid for Scientific Research from the Japan Society for Promotion of Science (Grants No. JP19H05826, JP22H00107).